\documentclass[pra,twocolumn,superscriptaddress,amsmath,amssymb,aps,floatfix]{revtex4-2}
\usepackage{graphicx}
\usepackage{dcolumn}
\usepackage{bm}
\usepackage{amsmath,amssymb}
\usepackage{graphics}
\usepackage{amsfonts}
\usepackage{epsfig}
\usepackage{float}
\usepackage{fancyhdr}
\usepackage{hyperref}
\usepackage{adjustbox}
\usepackage{float}
\usepackage{blindtext}
\usepackage{multirow}
\usepackage{graphicx}
\usepackage{dcolumn}
\usepackage{bm}
\usepackage{amsmath}
\usepackage{import}
\usepackage{physics}
\usepackage{verbatim}
\usepackage{listings}
\usepackage{xcolor}
\usepackage{xr-hyper}
\usepackage{hyperref}
\usepackage{cleveref}
\usepackage{newfloat}
\DeclareFloatingEnvironment[name={Fig S},fileext=lof]{suppfigure}
\begin{document}
\title{Generating highly entangled states via discrete-time quantum walks with Parrondo sequences}
\author{Dinesh Kumar Panda}
\email{dineshkumar.quantum@gmail.com}
\author{B. Varun Govind}
\email{bvarun.govind@niser.ac.in}
\author{Colin Benjamin}
\email{colin.nano@gmail.com}
\affiliation{School of Physical Sciences, National Institute of Science Education and Research Bhubaneswar, Jatni 752050, India}
\affiliation{Homi Bhabha National Institute, Training School Complex, Anushaktinagar, Mumbai
400094, India}
\begin{abstract}
Quantum entanglement has multiple applications in quantum information processing. Developing methods to generate highly entangled states independent of initial conditions is an essential task. Herein we aim to generate highly entangled states via discrete-time quantum walks. We propose deterministic Parrondo sequences that generate states that are generally much more entangled than states produced by sequences using only one of the two coins. We show that some Parrondo sequences generate highly entangled states, which are independent of the phase of the initial state used and further lead to maximally entangled states in some cases. We study Parrondo sequences for a small number of time steps and the asymptotic limit of a large number of time steps.
\end{abstract}
\maketitle
\section{\label{sec:level1}Introduction}

Parrondo's paradox involves two losing games, A and B, which, when combined, produce a winning outcome. These were introduced as a pedagogical illustration of the Brownian ratchet \cite{p1, parrondo_harmer_abbott_2000} and have spawned significant interest across diverse fields ever since \cite{Abbott2009,kanghao}. Parrondo's concept has been extended into other potential applications \cite{Abbott2009} and has been customized to explain many physical and biological processes, and their functioning \cite{allison_abbott_2001, bio1, bio2, bio3}. The quantum version of the paradox was introduced in Refs.~\cite{flitney2012quantum, meyer_2003}. Quantum walks were introduced in \cite{aharonov_davidovich_zagury_1993} and are a sophisticated tool for designing quantum algorithms \cite{portugal_2019}. A recent review \cite{review-qw-2021-d} deals with quantum walks and their application. Apart from uses in quantum computation, quantum walks have been used to explain, control, and simulate the dynamics in various complex physical systems \cite{portugal_2019}. Further, quantum walks can be directly implemented in labs without using a quantum computer \cite{portugal_2019}. Quantum walks have already been realized in ultra-cold Rubidium-$87$ atoms \cite{manouchehri_wang_2009}, photons \cite{card,wtf}, neutral atoms \cite{karski_forster_choi_steffen_alt_meschede_widera_2009}, NMR quantum-information processors \cite{robens_brakhane_meschede_alberti_2016}, superconducting qubits \cite{ryan_laforest_boileau_laflamme_2005, zhou_cai_su_yang_2019} and trapped ions \cite{flurin_ramasesh_hacohen-gourgy_martin_yao_siddiqi_2017, schmitz_matjeschk_schneider_glueckert_enderlein_huber_schaetz_2009}. Discrete-time quantum walks (DTQWs) allow for the exploration of numerous non-trivial geometric, and topological phenomena \cite{zzz}, and multi-path quantum interference \cite{puentes_2017}. An evolution by repeated application of two operators, the coin operator, and the shift operator, characterizes DTQWs. DTQW dynamics represents a useful tool in engineering arbitrary quantum states \cite{LI, gio}. The quantum-state transfer has been achieved with DTQW's using a position-dependent coin \cite{Kur}. On the other hand, continuous-time quantum walk (CTQW) is characterized by the free evolution of an $N$-dimensional quantum state under a Hamiltonian. It is represented by probability amplitudes assigned to each vertex in a graph on $N$ vertices. CTQWs have applications in search and sampling problems. Implementing logic gates via CTQW in dynamic graphs has been exhibited in Ref.~\cite{her}. Experimental investigation of CTQW's of single and correlated photons on discrete lattices has made huge progress due to the development of single-photon sources and the fabrication of coupled optical waveguides \cite{per}. Parrondo's paradox has been explored several times using one-dimensional DTQW's \cite{flitney2012quantum, meyer_2003,chandrashekar_banerjee_2011,neves_puentes_2018}. It was shown that instead of a single coin if a two-coin initial state is considered, Parrondo's paradox can be observed in quantum walks even in the asymptotic limits \cite{rajendran_benjamin_2018}, and further replacing a two-state coin(qubit) with a three-state coin(qutrit) also leads to Parrondo's paradox in asymptotic limits \cite{raj}.

Recently, a renewed interest in exploiting DTQWs as a tool to generate entanglement has been reported. In this context, developing faster methods to generate entanglement, which are also experimentally feasible, is an essential task of current research. A particular and intriguing way is via hybrid entanglement. As the name suggests, hybrid entanglement is the entanglement between different degrees of freedom like polarization, orbital angular momentum, time-bin energy, or spatial mode. Suppose these degrees of freedom belong to the same particle or qubit. In that case, it is called single-particle entanglement (SPE) or single-qubit entanglement(SQE), and more information can be encoded at the single-particle level, which can help to reduce the required resources \cite{gratsea_lewenstein_dauphin_2020}. Using SPE as a resource, a protocol to transfer the state of an unknown qubit to a distant location has been formulated \cite{pra}. SPE can be used to analyze states of photons, quantum liquids, and elementary particles~\cite{can}. An approach to manipulating SPE and quantum superposition with a single photon has the potential to be helpful in the engineering of single photon quantum devices~\cite{lu}. A physical process named 'quantum joining' and its inverse has been reported, which allows the transfer of entanglement between photons into hybrid entanglement of a single output photon leading to applications in quantum networking \cite{vit}. Single particle entangled states are relatively easier to produce than inter-particle entanglement and are also robust against decoherence and dephasing \cite{aqt}.
SPE has been used in some experimental tests of non‐contextual realistic hidden variable theories \cite{aqt}. Photonic hybrid entangled states can be advantageous in optical quantum networks as it enables a more flexible network with each photon being transmitted through the more suited channel~\cite{zhu_xiao_huo_xue_2020}. Photonic quantum information offers several technological applications \cite{slussarenko_pryde_2019}. Photonic implementations of DTQW's have been explored several times \cite{zhu_xiao_huo_xue_2020}. For a review on SPE, see Ref.~\cite{aqt}. In this scenario, preparing maximally hybrid entangled states using DTQWs is of great importance, as it would give rise to new possibilities concerning quantum technologies. Hybrid entanglement generation has been realized experimentally as well~\cite{li_yan_he_wang_2018}.

Here, we discuss the generation of hybrid entangled states generated in DTQWs. Several approaches have been explored to enhance the entanglement between the position and coin degrees of freedom of a single qubit walker. A methodical study of the entanglement properties of disordered quantum walks was carried out in Ref.~\cite{ch2012disorder}. It was shown that randomly choosing the coin operator at each time step of the quantum walk can lead to maximally entangled states in the asymptotic limit independent of the initial state, but only one initial state was analyzed. The effect of the introduction of different types of disorder in the generation of highly entangled states was further studied using several numerical experiments~\cite{vieira_amorim_rigolin_2013, vieira_amorim_rigolin_2014}. However, this strategy requires many steps to reach the asymptotic limit, which is disadvantageous for current experiments. Gratsea et al. suggested the optimization of the coin operator sequence using a basin-hopping algorithm \cite{gratsea_lewenstein_dauphin_2020}. It was shown that maximal entanglement is achieved in just $10$ steps. The small number of steps meant that the experimental challenge of reaching the asymptotic limit was solved. However, the experimental implementation potentially requires a complete set of possible coin operators. Recently, two new approaches were proposed in \cite{gratsea2020universal}. The first approach presented an entangling sequence (Universal entangler) consisting of a deterministic sequence of Hadamard and Fourier coin operators that created a universal amount of entanglement for a class of localized initial states with vanishing relative phase. This sequence had the disadvantage of only working for an odd number of steps. The second method was based on direct optimization of the coin sequence using reinforcement learning (RL), which is a technique that allows us to determine longer sequences (Optimal) where brute force optimization is not possible. Although it was shown that optimal sequences were better entanglers than universal entanglers, they were highly dependent on initial states.

Our main motivation in this work is to exploit the strategy of turning two unfavorable situations into a favorable one to generate hybrid highly entangled states in DTQWs. We propose using deterministic Parrondo sequences using two different coin operators to produce states that are generally more entangled than the states produced by sequences using only one of the two coins. We show that some of the deterministic sequences proposed, apart from generating highly entangled states, generate entanglement independent of the initial state phase used $(\text{phase independent entanglers}: XXH..., XXM..., XXF...)$. We also show that phase-independent entangling sequences generate maximally entangled states for $3$ and $5$ steps regardless of initial states. It could resolve the experimental challenges of generating these states. Further, we explore the asymptotic limit of the sequences proposed. Recently, a few studies have analyzed the relationship between Parrondo's games and quantum entanglement in DTQW's, see \cite{jan2020study, PhysRevA.97.012116, wal, pir}. In \cite{jan2020study}, winning and losing Parrondo games are defined as in \cite{rajendran_benjamin_2018}, and the winning Parrondo's games are shown to generate a good amount of entanglement. However, only two initial states are explored ($\theta = 0, \phi = \pi/2$ and $\theta = 0, \phi = 3\pi/2$), and the sequences which possess maximum winning outcomes are not the best entanglers.
Ref. \cite{PhysRevA.97.012116} presents the effect of periodic sequence on the entanglement between coin and position space. Two different coins are considered, showing that the periodic introduction of the second coin to the quantum walk determined only by the first coin increases entanglement. However, the quantum coin under consideration is a single non-orthogonal, unitary operator, and the results presented are only for a particular initial state. Ref.~\cite{wal} also defines winning and losing Parrondo sequences as in \cite{rajendran_benjamin_2018} and further shows, with some examples of coin operators acting on a particular initial state, that Parrondo's effect can also occur in the case of 1D DTQW with a deterministic aperiodic sequence of two-state quantum coins. With a period-one sequence of the coin operators, the walker-coin quantum entanglement is lesser, on average, than in the case of quantum walks with deterministic aperiodic sequences. Ref.~\cite{pir} exhibits similar behavior where time-dependent quantum coins have been applied.

The paper is organized as follows: In the theory section, we briefly review DTQWs and discuss how the Schmidt norm can be used to measure entanglement. We then introduce Parrondo sequences. In the results section, we present and analyze various Parrondo sequences as regards their propensity to generate highly entangled states and phase-independent hybrid entanglement. We then discuss the results and compare some aspects with other relevant works in the discussion section. Finally, in the conclusion section, we summarize our work and future goals. In supplementary material, we have discussed some additional works and provided our computational python code.

\section{Theory}
\subsection{Discrete time quantum walks (DTQW's)}
DTQW's in 1D are defined on a tensor product of two Hilbert spaces $H = H_P \otimes H_C $. $H_C$ represents the 2D Hilbert space associated with the coin or qubit, whose computational basis is $\{\ket{0_c},\ket{1_c}\}$ and $H_P$ represents the infinite-dimensional space associated with the position of the walker, whose computational basis is $\{\ket{n_p}: n_p \in Z\}$. The walker is assumed to be initially localized at position `$0$' in a general superposition of coin states,
\begin{equation}
\ket{\psi_0} = \cos(\theta/2)\ket{0_p0_c} + e^{i\phi}\sin(\theta/2)\ket{0_p1_c},
\label{3.1}
\end{equation}
\noindent
where $\theta \in [0, \pi]$ and $\phi \in [0, 2\pi]$. A unitary operator describes the shift, called the shift operator
\begin{align}
\begin{split}
S = & \sum_{j=-\infty}^{\infty}(\ket{(j+1)_p}\bra{j_p}\otimes\ket{0_c}\bra{1_c}+\ket{(j-1)_p}\bra{j_p}\otimes\ket{1_c}\bra{0_c}),
\label{3.2}
\end{split}
\end{align}
\noindent
and the unitary coin operator is
\begin{equation}
C(\alpha, \beta, \gamma, \eta) = e^{i\eta/2}
\begin{pmatrix}
e^{i\alpha}\cos\beta & e^{i\gamma}\sin\beta\\
-e^{-i\gamma}\sin\beta & e^{-i\alpha}\cos\beta

\end{pmatrix}.
\label{coin}
\end{equation}
\noindent
The full evolution then can then be described as
\begin{equation}
U(t) = S.C(\alpha(t), \beta(t), \gamma(t), \eta(t)).
\label{3.8}
\end{equation}
\noindent
The time evolution of the system after $t$ steps is then
\begin{align}
\begin{split}
\ket{\psi_t} & = U(t)\ket{\psi_{t-1}} = U(t)U(t-1)...U(1) \ket{\psi_0}\\
& =\sum_{j=-\infty}^{\infty}[a_{1}(j,t)\ket{j_p,0_c} + a_{2}(j,t)\ket{j_p,1_c}]\;,
\label{3.9}
\end{split}
\end{align}
\noindent
where, $a_{1}(j,t)\text{ and } a_{2}(j,t)$ are amplitudes of states: $\ket{j_p,0_c}$ and $\ket{j_p,1_c}$ respectively.
\noindent
\subsection{Measure of Entanglement}
Entanglement measures quantify the amount of entanglement in a quantum state. There are several measures of entanglement like concurrence\cite{hildebrand_2007}, logarithmic negativity\cite{plenio_2005}, and entropy of entanglement\cite{janzing_2009}. Following Ref. \cite{gratsea_lewenstein_dauphin_2020}, we use Schmidt Norm as a measure of entanglement, which is relatively simpler to calculate for quantum walks since it only deals with pure states. In this work, the initial state is always pure, and the state evolves unitarily. Thus it remains a pure state at any later time too. This fact allows us to write the density operator $\rho$ of $\ket{\psi}$ directly as $\rho = \ket{\psi}\bra{\psi}$. As a consequence of singular value decomposition, the normalized state $\ket{\psi}$ can be re-expressed in a Schmidt-decomposed form\cite{reuvers_2018},
\begin{equation}
\ket{\psi} = \sum_{i = 1}^{k} \lambda_{i} \ket{\psi_i}_P \otimes \ket{\psi_i}_C,
\label{9}
\end{equation}
\noindent
with orthonormal Schmidt vectors $\ket{\psi_i}_P \in H_P$, $\ket{\psi_i}_C \in H_C$ and real non-negative Schmidt coefficients $\lambda_{i}$ satisfying the normalization condition $\sum_{i = 1}^{k} |\lambda_{i}|^2 = 1$ with $\lambda_1 \geq \lambda_2 \geq....\geq \lambda_k \geq 0$ where $k = min(d_P,d_C)$ is written in terms of dimensions $d_P$ and $d_C$ of the position and coin space. The reduced density operator for the coin space is defined by $\rho_C \equiv \text{Tr}_p(\rho)$ where $\text{Tr}_p$ is the partial trace over the position space. It gives,
\begin{align}
\begin{split}
\rho_{C} = & \alpha(t)\ket{0}\bra{0} + \beta(t)\ket{1}\bra{1} + \gamma(t)\ket{0}\bra{1} + \gamma^*(t)\ket{1}\bra{0},
\end{split}
\label{3.16}
\end{align}
\noindent
wherein $\alpha(t) = \sum_{j=-\infty}^{\infty}|a_{1}(j,t)|^2$, $\beta(t) = \sum_{j=-\infty}^{\infty}|a_{2}(j,t)|^2$ and $\gamma(t) = \sum_{j=-\infty}^{\infty} a_{1}(j,t)a_{2}^*(j,t)$. The Schmidt norm\cite{reuvers_2018} is defined as,
\begin{equation}
||\psi||_{z,k} = \Big(\sum_{i=1}^{k} (\lambda_i)^z\Big)^{1/z}\;.
\label{3.17}
\end{equation}
\noindent
Setting $z = 1$, and for the system considered, $k = min(d_P,d_C) = 2$. The Schmidt norm for a maximally entangled state is $\sqrt{2}$, since $\lambda_i = 1/\sqrt{k} = 1/\sqrt{2}$. Schmidt norm is calculated analytically using the relation between Schmidt coefficients and eigenvalues of the reduced density matrix, see \cite{gratsea_lewenstein_dauphin_2020}. The reduced density matrix is given as,
\begin{equation}
\rho_C = \Tr_p(\rho) = \frac{I}{2} + \vec{n}\cdot\vec{\sigma}\;.
\label{trace-dkp9}
\end{equation}
\noindent
The trace in Eq.~(\ref{trace-dkp9}) is taken over position degrees of freedom. $I$ is the identity matrix, $\vec{\sigma}$ is the Pauli vector and,
\begin{align}
\begin{split}
\vec{n} = & \Big(\Re(\Sigma_j a_{1}(j,t)a_{2}^*(j,t)), \Im(\Sigma_j a_{1}(j,t)a_{2}^*(j,t)),\\ & \frac{1}{2}\Sigma_j(|a_{1}(j,t)|^2 - |a_{2}(j,t)|^2)\Big).
\label{3.23}
\end{split}
\end{align}
\noindent
The eigenvalues of the reduced density matrix($\rho_{C}$) are then given by,
\begin{align}
\begin{split}
E_{\pm} & = \frac{1}{2} \pm \sqrt{\frac{1}{4} - \alpha(t)(1 - \alpha(t)) + |\gamma(t)|^2}= \frac{1}{2} \pm |\vec{n}|\;.
\label{3.26}
\end{split}
\end{align}
\noindent
From the identity $E_{\pm} = \lambda^{2}_{\pm}$, one can rewrite the Schmidt norm as,
\begin{equation}
S = ||\psi||_{1,2} = \sqrt{E_-} + \sqrt{E_+} \;.
\label{3.28}
\end{equation}
\subsection{Parrondo sequences}
Parrondo's paradox states that one can strategize to turn two unfavorable strategies into a favorable one by combining them in a random or deterministic way. We can categorize Parrondo sequences in DTQW's into three categories: (1) Deterministic sequences- the coin used at each step is wholly specified before starting the sequence of coins. (2) Random sequences- the coin used at each step is decided randomly. (3) Adaptive sequences- the coin to be used at each time step is decided based on some criterion as the walk proceeds. Our motivation is to generate highly hybrid entangled states using deterministic Parrondo sequences. We are interested in the small number of time steps and the asymptotic setting of a large number of time steps.
\section{Results}
We restrict ourselves to a set of coin operators, Eq.~(\ref{coin}): the Hadamard coin $H=C(\alpha=-\pi/2, \beta =\pi/4, \gamma=-\pi/2, \eta=\pi)$, the Fourier coin $F=C(\alpha=0, \beta=\pi/4, \gamma=\pi/2, \eta=0)$, the Miracle coin $M=C(\alpha=\pi/2, \beta=\pi/4, \gamma=0, \eta=0)$, the flip or Grover coin $X=C(\alpha \in [0,\pi/2], \beta=\pi/2, \gamma=-\pi/2, \eta=\pi)$, which have the explicit forms:
\begin{align}
\begin{split}
H & = \frac{1}{\sqrt{2}}
\begin{pmatrix}
1 & 1 \\
1 & -1
\end{pmatrix}, \hspace{0.6cm}
F = \frac{1}{\sqrt{2}}
\begin{pmatrix}
1 & i \\
i & 1
\end{pmatrix}, \\
M & = \frac{1}{\sqrt{2}}
\begin{pmatrix}
i & 1 \\
-1 & -i
\end{pmatrix}, \hspace{0.4cm}
X =
\begin{pmatrix}
0 & 1 \\
1 & 0
\end{pmatrix}.
\end{split}
\end{align}
\noindent
We performed several numerical experiments with these coins by forming various deterministic Parrondo sequences like $ABAB..., BAABAA..., AABAAB...$, etc., where $A, B \in\{H, F, M, X\}$. The primary idea was to find sequences involving two coins that could perform better than sequences formed using only either coin ($AAA...$ and $BBB...$). We narrowed it down to a few better-performing sequences based on the experiments. {The best result of our work is presented here, and the rest are shown in the supplementary section. However, the results from the supplementary section are used in the discussion section.}
\begin{figure*}[t!]
{ \begin{minipage}[b]{8.6 cm}
\includegraphics[width = 8.6 cm]{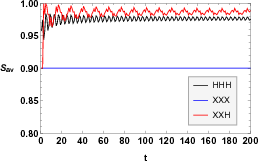}
\text{(a) $XXH...$}
\end{minipage}}
\hfill
{ \begin{minipage}[b]{8.6 cm}
\includegraphics[width = 8.6 cm]{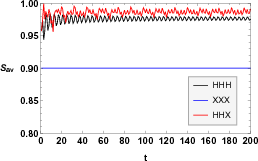}
\text{(b) $HHX...$}
\end{minipage}}
\hfill
{ \begin{minipage}[b]{8.6 cm}
\includegraphics[width = 8.6 cm]{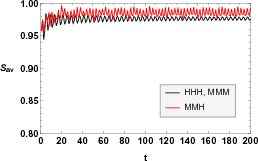}
\text{(c) $MMH...$}
\end{minipage}}
\hfill
{ \begin{minipage}[b]{8.6 cm}
\includegraphics[width = 8.6 cm]{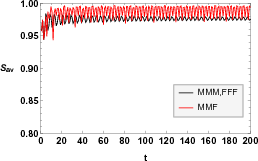}
\text{(d) $MMF...$}
\end{minipage}}
\caption{Average Schmidt norm $S_{av}$ from Eq.~(\ref{eq-svalue}) after evolution with some deterministic Parrondo sequences: (a) $XXH...$ for $t = 200$ steps compared with $HHH...$, and $XXX...$ . The plots for $XXF...$ and $XXM...$ produced similar results. (b) $HHX...$ for $200$ time steps compared again with $HHH...$ and $XXX...$ . (c) $MMH...$ for $200$ time steps compared with $MMM...$ and $HHH...$ . (d) $MMF...$ for $200$ time steps compared with $MMM...$ and $FFF...$ .}
\label{fig0}
\end{figure*}
\begin{figure*}[t!]
\begin{minipage}[b]{8.6 cm}
\includegraphics[width = 8.6 cm]{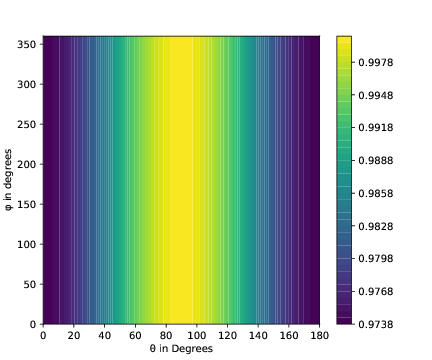}
\text{(a) $XXH...$}
\end{minipage}
\begin{minipage}[b]{8.6 cm}
\includegraphics[width = 8.6 cm]{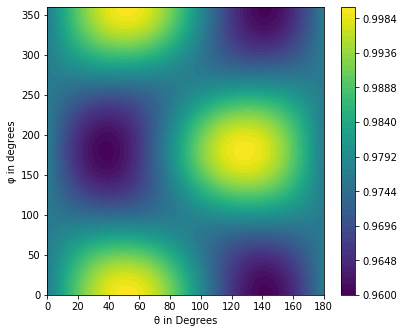}
\text{(b) $HHH...$}
\end{minipage}
\hfill
\begin{minipage}[b]{8.6 cm}
\includegraphics[width = 8.6 cm]{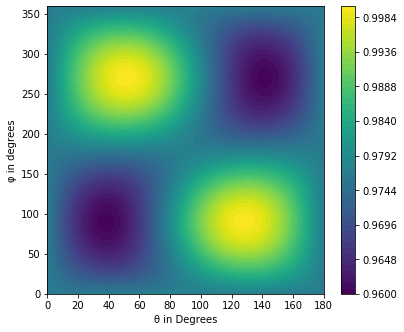}
\text{(c) $FFF...$}
\end{minipage}
\hfill
\begin{minipage}[b]{8.6 cm}
\includegraphics[width = 8.6 cm]{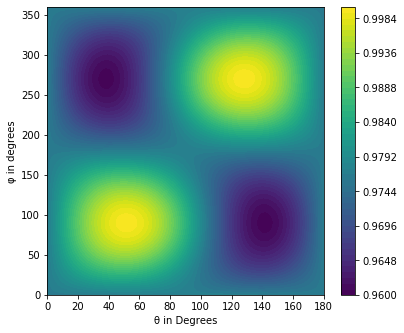}
\text{(d) $MMM...$}
\end{minipage}
\caption{The contour plots of Schmidt norm $(S)$ shown are at time step $t=50$ and are generated for $180$ different $\theta$ values and $360$ different $\phi$ values giving a total of $64800$ different initial states. (a) Contour plot of Schmidt norm as a function of the initial state parameters $\phi$ and $\theta$ for $XXH...$ for a $50$ time step DTQW. The contour plots for $XXF...$ and $XXM...$ were identical to that of $XXH...$ . Contour plots of Schmidt norm as a function of the initial state parameters $\phi$ and $\theta$ for (b) $HHH...$ (c) $FFF...$ and (d) $MMM...$ for $50$ time steps.}
\label{fig1}
\end{figure*}
\subsection{Parrondo Sequence-$AABAAB...$}
The first set of sequences we discuss is of the type $AABAAB...$ . We first focus on cases where $A$ is always the flip coin $X$ and $B$ can be Hadamard coin $H$, Fourier coin $F$, or Miracle coin $M$. Fig.~\ref{fig0}(a) shows the plot of Schmidt norm $S$ as a function of number of time steps $t$ after evolution with the sequences $XXH..., HHH...$ and $XXX...$ for $t=200$ time steps.
Each point in this plot is an average $S$ taken over the whole range of initial states, i.e., the complete range of $\theta$ and $\phi$ and is evaluated using the following integral:
\begin{equation}
S_{av}=\langle\frac{S}{\sqrt{2}}\rangle = \frac{1}{2\pi^{2}}\int_{0}^{2\pi}d\phi \int_{0}^{\pi}d\theta \;  \; \frac{S}{\sqrt{2}} \;\;\;.
\label{eq-svalue}
\end{equation}
We use Eq.~(\ref{eq-svalue}) to evaluate average S in the entire manuscript. The existence of Parrondo strategies is evident from Figure~\ref{fig0} for a small number of time steps and in the asymptotic setting. Sequences $XXF...$ and $XXM...$ produced results identical to $XXH...$ . This fact is verified from the equivalence of the analytical expressions of the Schmidt norm in terms of the initial state parameters $\phi$ and $\theta$ for all $XXH..., XXF...$ and $XXM...$ . For the first six steps, the expressions for Schmidt norm $(S)$ are given as follows:
\begin{align}
\begin{split}
\text{After} \, 1 \, \text{step}\xrightarrow[]{}&\frac{1}{\sqrt{2}}\left(\sqrt{1-\cos (\theta)}+\sqrt{1 + \cos (\theta)}\;\right),\\
\text{After} \, 2 \, \text{steps}\xrightarrow[]{} &\frac{1}{\sqrt{2}}\left(\sqrt{1-\cos (\theta)}+\sqrt{1 + \cos (\theta)}\;\right),\\
\text{After} \, 3 \, \text{steps}\xrightarrow[]{} & \sqrt{2},\\
\text{After} \, 4 \, \text{steps}\xrightarrow[]{}&\frac{1}{2} \left(\sqrt{2 + \sin (\theta)}+\sqrt{2- \sin (\theta)}\;\right),\\
\text{After} \, 5 \, \text{steps}\xrightarrow[]{} & \sqrt{2},\\\
\text{After} \, 6 \, \text{steps}\xrightarrow[]{} &\frac{1}{2\sqrt{2}}\left( \sqrt{4 + \sin (\theta)}+ \sqrt{4- \sin (\theta)}\;\right).
\end{split}
\end{align}
The analytical expressions reveal that the value of entanglement $S$ is independent of the phase $\phi$ at each step. Due to this property, we refer to the sequences $XXH..., XXF...$ and $XXM...$ as phase-independent entanglers. It is seen from the above expressions that the phase-independent entanglers generate maximally entangled states for $t=3$ and $t=5$ step DTQW's for all initial states. We propose some more Parrondo sequences belonging to the $AABAAB...$ type. Fig.~\ref{fig0}(b-d) show the plots of Schmidt norm $S_{av}$ as a function of number of time steps $t$ after evolution with sequences $HHX...$, $MMH...$ and $MMF...$ for $200$ time steps. It has to be noted that the results of Parrondo sequences $FFX..., HHM..., FFM...$ are the same as $HHX..., MMH..., MMF...$ respectively, see Eq.~(\ref{eq-svalue}). It can be attributed to the fact that $F$ and $M$ act as phase shifted $H$, and is discussed below. Also, it is to be noted that the above sequences are not phase-independent entanglers.

Figure~\ref{fig1}(a) shows the Schmidt norm as a function of the initial state parameters $\theta$ and $\phi$ for the phase independent entanglers, verifying the independence of phase $\phi$ in the value of entanglement $S$. In order to gain further insight, the plots of Schmidt norm as a function of the initial state parameters $\theta$ and $\phi$ for sequences $HHH..., FFF... $ and $MMM...$ are shown in Fig.~\ref{fig1}(b-d). A simple comparison between the Schmidt norm plots for $HHH..., FFF...$ and $MMM...$ tells us that the sequences $FFF...$ and $MMM...$ generate values of $S$ that are shifted by a phase of $-\pi/2$ and $+\pi/2$ respectively concerning the value of $S$ generated by $HHH...$ . This fact can be further verified by comparing the analytical expressions of $S$ as a function of $\theta$ and $\phi$ for the sequences $HHH..., FFF...$ and $MMM...$ . For example, consider the analytical expressions of $S$ in the form $\sqrt{E_-} + \sqrt{E_+}$ for a single step DTQW evolved using $H$, or $F$ or $M$,
\begin{align}
\begin{split}
H\xrightarrow[]{}&\frac{1}{\sqrt{2}}\left(\sqrt{1+\sin(\theta)\cos(\phi)}+ \sqrt{1-\sin(\theta)\cos(\phi) }\;\right)\\
F\xrightarrow[]{}&\frac{1}{\sqrt{2}}\left(\sqrt{1+\sin(\theta)\sin(\phi)}+ \sqrt{1-\sin(\theta)\sin(\phi)}\;\right)\\
M\xrightarrow[]{}&\frac{1}{\sqrt{2}} \left(\sqrt{1 - \sin (\theta) \sin (\phi)} + \sqrt{1 + \sin (\theta) \sin (\phi) }\;\right)
\end{split}
\end{align}
Substituting the identities $\cos(\phi - \pi/2) = \sin(\phi)$ and $\cos(\phi + \pi/2) = -\sin(\phi)$ in the analytical expression of $S$ for $H$ verifies this assertion. The same fact can be verified for any time step $t$. Hence, the equivalence in results between $XXH..., XXF...$ and $XXM...$ can be attributed to the fact that the Fourier coin $F$ and the miracle coin $M$ act like phase-shifted Hadamard coin $H$ when considered in terms of entanglement generation. {The transport properties of the walker can also be explored using various Parrondo sequences discussed and the study in higher spatial dimensions\cite{spe-2d-franco} is a future goal. }

\section{Discussion}
\begin{figure*}[t!]
\begin{minipage}[b]{14.5 cm}
\includegraphics[width = 14 cm]{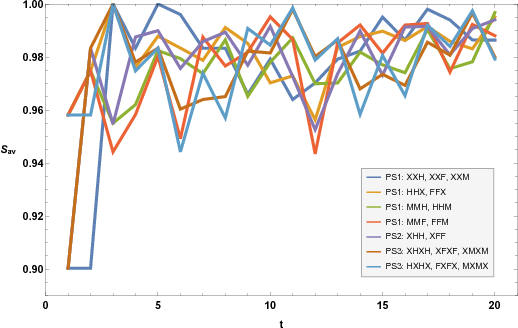}

\end{minipage}
\caption{Comparison between various Parrondo sequences: $PS1=AABAAB...,\; PS2=ABBABB...,\; PS3=ABAB...$, with $A,B \in \{H, F, M, X\}$ for small number of time steps(t). Each point is an average Schmidt norm ($S_{av}$) value based on Eq.~(\ref{eq-svalue}).}
\label{fig5}
\end{figure*}

We now compare the various Parrondo sequences suggested in Figure~\ref{fig5}. It has already been seen that when each point is averaged for {all or} a large number of states, the plots for some of the cases within a specific sequence produce identical results. {However, it has to be noted that identical results are not seen when one particular state is considered in some sequences. For example, MMH.. and HHM.. do not give identical results for a particular state. In such cases, the phase $\phi$ of the initial state will account for the difference in entanglement values. When {all or} a large number of states are considered and averaged, the role of the phase of the initial state will be averaged out.}
Sequences with identical results have been grouped in Fig.~\ref{fig5}. Fig.~\ref{fig5} suggests that for any small time step, we could find coin sequences that yield high average values of entanglement, which is particularly useful in cases where the initial state is not fully known, very noisy, or whenever it is required to generate highly entangled states independent of the initial state \cite{gratsea2020universal}.
\begin{figure}[h!]
\includegraphics[width = 8.5 cm]{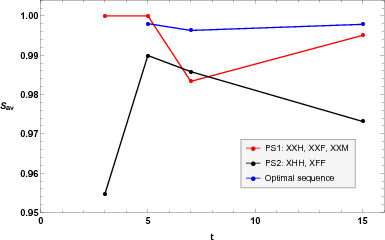}
\caption{Parrondo sequences(PS1, PS2) vs. Optimal sequences(see Gratsea et al. \cite{gratsea2020universal}). Average Schmidt norm $S_{av}$ is plotted for $3,\;5,\;7,\;15$ step quantum walks.}
\label{fig7}
\end{figure}

\makeatletter
\setlength{\@fptop}{0pt}
\makeatother
\begin{table*}[t]
\centering
\caption{Comparison between various approaches}
\label{tab7}
\resizebox{\textwidth}{!}{%
\begin{tabular}{|c|l|l|l|l|}
\hline
\textbf{Properties$\downarrow$/Model$\rightarrow$} &
\multicolumn{1}{c|}{\textbf{\begin{tabular}[c]{@{}c@{}} DTQW with Parrondo sequences\\ (This paper) \end{tabular}}} &
\multicolumn{1}{c|}{\textbf{\begin{tabular}[c]{@{}c@{}} DTQW with disorder\\ Refs.~\cite{vieira_amorim_rigolin_2013,vieira_amorim_rigolin_2014}\end{tabular}}} &
\multicolumn{1}{c|}{\textbf{\begin{tabular}[c]{@{}c@{}}DTQW with Optimization\\ (basin hopping algorithm) \\ Ref.~\cite{gratsea_lewenstein_dauphin_2020}\end{tabular}}} &
\multicolumn{1}{c|}{\textbf{\begin{tabular}[c]{@{}c@{}}DTQW with Optimization\\(Reinforcement Learning technique)\\ Ref.~\cite{gratsea2020universal}\end{tabular}}} \\ \hline
\hline
\textbf{\begin{tabular}[c]{@{}c@{}}Dependence\\ of Entanglement\\ on initial state \\ parameters\end{tabular}} &
\begin{tabular}[c]{@{}l@{}} $S$ is independent of phase $\phi$ \\ for $XXH..., XXM...$ and $XXF...$ \end{tabular} &
\begin{tabular}[c]{@{}l@{}}Asymptotic limit \\ was shown to be \\ independent of\\ initial parameters\end{tabular} &
\begin{tabular}[c]{@{}l@{}}$\theta$ and $\phi$ \\ dependent\end{tabular} &
\begin{tabular}[c]{@{}l@{}}Universal Entangler: \\ $\theta$ independent for $\phi = 0$ states\\ Optimal sequence: \\ $\theta$ and $\phi$ dependent\end{tabular} \\ \hline
\textbf{\begin{tabular}[c]{@{}c@{}}No of \\ coins used\end{tabular}} &
Two coins per sequence &
\begin{tabular}[c]{@{}l@{}}2 coins in $RQRW_2$,\\ full set of possible coins\\ in $RQRW_\infty$\end{tabular} &
\begin{tabular}[c]{@{}l@{}}Full set of \\ possible coins\end{tabular} &
2 coins \\ \hline
\textbf{5 steps} &
\begin{tabular}[c]{@{}l@{}}Maximal Entanglement. \end{tabular} &
\begin{tabular}[c]{@{}l@{}} Not discussed. \end{tabular} &
\begin{tabular}[c]{@{}l@{}}Around $98\%$ entanglement. \end{tabular} &
\begin{tabular}[c]{@{}l@{}}Optimal sequence: $0.998$\\ \end{tabular} \\ \hline
\textbf{20 steps} &
\begin{tabular}[c]{@{}l@{}}Maximum average value\\ for MMH..., HHM...: $S_{av} = 0.99675$\\ See Fig.~\ref{fig5}\end{tabular} &
\begin{tabular}[c]{@{}l@{}} Not discussed \end{tabular}&
\begin{tabular}[c]{@{}l@{}} Not discussed\end{tabular} &
\begin{tabular}[c]{@{}l@{}} Not discussed\end{tabular}\\ \hline
\end{tabular}%
}
\label{tab}
\end{table*}
Finally, we compare our results with the previous works~\cite{vieira_amorim_rigolin_2013, vieira_amorim_rigolin_2014, gratsea2020universal, gratsea_lewenstein_dauphin_2020}. Fig.~\ref{fig7} shows a comparison plot between Parrondo strategies and optimal entanglers as in~\cite{gratsea2020universal}. Schmidt norm $S_{av}$ was plotted for $3, 5, 7, 15$ step DTQW's. Each data point was obtained based on Eq.~(\ref{eq-svalue}). The plot shows that Parrondo strategies $XXH..., XXF...$ and $XXM...$ produce higher values of average entanglement ($S_{av}$) than the optimal entanglers for $3, 5$ step quantum walks. Also, $XXH..., XXF..., \text{and } XXM...$ give maximally entangled states ($S_{av} = 1$) regardless of the initial state used for $3$ and $5$ step quantum walks. It is valuable for an experimental setting where the number of possible steps should be as small as possible. It serves as an example for cases where we should find coin sequences that yield high average values of entanglement over a large number of initial states for a given $t-$ time step quantum walk. It is important in an experimental setting where the initial state/state is not fully known~\cite{gratsea2020universal}. {We} compare our best results with all the other works \cite{vieira_amorim_rigolin_2013, vieira_amorim_rigolin_2014, gratsea2020universal, gratsea_lewenstein_dauphin_2020} discussed focusing on the similarities and differences among them in Table~\ref{tab}. The first comparison we make is of the type of sequences used. We used deterministic sequences instead of choices made by Vieira et al. ~\cite{vieira_amorim_rigolin_2013, vieira_amorim_rigolin_2014} where disordered sequences are used. Gratsea et al. mainly focus on optimization to find sequences but also propose one deterministic sequence - Universal entangler. The independence of entanglement $S$ on initial state parameter $\phi$ for all states and for any time step is a unique feature seen only in $XXH...,\; XXF...,\; XXM...$ Parrondo sequences. For initial states with $\phi = 0$, $\theta$ independence is observed for the universal entanglers at any time step. In the asymptotic limit, disordered sequences discussed by Vieira et al. produced maximal entanglement irrespective of the initial state parameters. Another important property to be considered is the number of coin operators used. Lesser the number of coins, the easier the experimental realization. In terms of this property, Parrondo sequences are as good as the universal and optimal entanglers \cite{gratsea2020universal}, and $RQRW_2$\cite{vieira_amorim_rigolin_2013, vieira_amorim_rigolin_2014}. Gratsea et al. \cite{gratsea2020universal} focus only on a small number of time steps. In their paper \cite{gratsea_lewenstein_dauphin_2020}, it was shown that maximal entanglement is achieved in just $10$ steps. Optimal entangler generates an entanglement value of around $0.998$ for $5-$time steps, but Parrondo sequences $XXH..., XXF..., XXM...$ give maximal entanglement for both three and five-time steps. The case of small number of time steps is not discussed in Refs.~\cite{vieira_amorim_rigolin_2013, vieira_amorim_rigolin_2014}. In the asymptotic limit, Vieira et al. \cite{vieira_amorim_rigolin_2013,vieira_amorim_rigolin_2014} showed that disordered sequences achieve maximal entanglement.

\section{Conclusions and Future goals}
The central goal of this work was to show that deterministic Parrondo strategies can generate highly entangled states when implemented on DTQWs. Using the Schmidt norm as the measure of entanglement, we showed that several deterministic sequences formed using two coin operators were better entanglers than single coin sequences made of either one of the two coins. We also showed that the Parrondo sequences $XXH..., \; XXF...$ and $XXM...,$ generate entanglement independent of the initial state phase. Furthermore, we showed $XXF..., \; XXH...$ and $XXM...,$ generate maximally entangled states for $3$ and $5$ step quantum walks for all initial states. A comparison between various Parrondo sequences showed that for the average Schmidt norm, the plots of some of the sequences within a type of strategy (like $AABAAB...,\; ABAB...$) produce similar results. The similarities in results were understood to be associated with the fact that the Fourier coin $F$ and the miracle coin $M$ act like phase-shifted Hadamard coin $H$. The comparison also suggested that, for any number of time steps $t$, one could find coin sequences that yield high average values of entanglement. It is advantageous in cases where the initial state is not fully known, is very noisy, or when it is required to generate highly entangled states independent of the initial state \cite{gratsea2020universal}. It was also seen that the strategies displayed Parrondo's paradox even when a large number of steps are considered.

In the future, we plan to extend our work to 2D and look at how entanglement between X and Y position degrees of freedom for the walker evolve among themselves and separately with coin degree of freedom. There is the possibility of generating delocalized and localized probability distributions for the quantum walker(particle) evolving via a quantum walk with different Parrondo sequences(AAA..., BBB..., ABAB..., AAB..., ABB... and so on). So exploring possible connections between the spread of the walker with the entanglement generated for various Parrondo sequences could be a topic of future research. Of course, this is a challenge for experimentalists to explore and implement our proposed method.
\section{Acknowledgments}
Colin Benjamin would like to thank Science and Engineering Research Board (SERB) for funding under MATRICS grant "Nash equilibrium versus Pareto optimality in N-Player games" (MTR/2018/000070) and Core Research grant "Josephson junctions with strained Dirac materials and their application in quantum information processing," Grant No. CRG/2019/006258.

\twocolumngrid

\newpage
\widetext

\section{Supplementary Material}
In this supplementary material, subsections A and B present some more Parrondo sequences, which result in highly entangled states. In subsection C, we provide the code needed to generate a typical paper plot.

\subsection{Parrondo sequence- $ABBABB...$}

\begin{suppfigure}[h!]
\centering
\includegraphics[width = 9.0 cm]{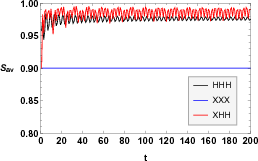}
\caption{Average Schmidt norm $S_{av}$ after evolution with the deterministic Parrondo sequences: $XHH...$ compared with $HHH...$ and $XXX...$ for $200$ time steps(t). The sequence $XFF...$ was found to produce similar results as $XHH...$ .}
\label{fig3}
\end{suppfigure}
Fig. S~\ref{fig3} shows the plots of average Schmidt norm $S_{av}$ as a function of number of time steps $t$ after evolution with the sequence $XHH...$ . Here again, a good enhancement in entanglement for $XHH...$ is seen compared to $HHH...$ and $XXX...$ . A similarity between the results of $XHH...$ and $XFF...$ was also found here. However, the results of $XMM...$ are not as good as $XHH...$ and $XFF...$ . The exact reason requires further study.

\subsection{Parrondo sequence- $ABAB...$}

\begin{suppfigure*}[h!]
\begin{minipage}[b]{8.0 cm}
\includegraphics[width = 8.0 cm]{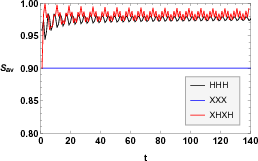}
\text{(a) $XHXH...$}
\end{minipage}
\hspace{0.3cm}
\begin{minipage}[b]{8.0 cm}
\includegraphics[width = 8.0 cm]{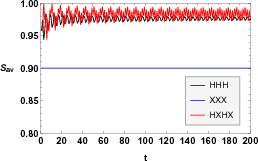}
\text{(b) $HXHX...$}
\end{minipage}
\hfill
\caption{(a) Average Schmidt norm $S_{av}$ after evolution with the deterministic Parrondo sequences $XHXH...$ versus $HHH...$ and $XXX...$ for $140$ time steps. (b) Average Schmidt norm $S_{av}$ after evolution with deterministic sequences $HXHX...$ versus $HHH...$ and $XXX...$ for $200$ time steps.}
\label{fig4}
\end{suppfigure*}

Fig. S~\ref{fig4} shows the plots of average Schmidt norm $S_{av}$ as a function of the number of time steps $t$ after evolution with sequences $XHXH...\text{ and}\; HXHX...$ . Sequences $XHXH...,\; XFXF...$ and $XMXM...$ generate similar results while sequences $HXHX...,\; FXFX...$ and $MXMX...$ generate similar results. As in the previous sections, this can be reasoned as a manifestation of the relation between $H, F$, and $M$.


\subsection{ Python code to generate the plots of Schmidt norm for Parrondo sequences }

The following python code is written by us based on \cite{quan} and gives the data points for the sequence $XXH...$ in Figure~\ref{fig0}(a) of the main manuscript.
{\onecolumngrid
\begin{scriptsize}
\begin{verbatim}
#XXH SEQUENCE:
from numpy import *
import numpy as np
import random
from scipy import integrate

N = 200 #Number of random steps
DKP_XXH = np.zeros(N)

#Needed stuff:
coin0 = np.array([1, 0]) # |0>
coin1 = np.array([0, 1]) # |1>
C00 = np.outer(coin0, coin0) # |0><0|
C01 = np.outer(coin0, coin1) # |0><1|
C10 = np.outer(coin1, coin0) # |1><0|
C11 = np.outer(coin1, coin1) # |1><1|

C_hat = (cos(pi/4)*C00 + sin(pi/4)*C01 + sin(pi/4)*C10 - cos(pi/4)*C11) #Hadamard coin
K_hat = (C01 + C10) #Grover coin
F_hat = (C00 + 1j*C01 + 1j*C10 + C11)/sqrt(2.) #Fourier coin
M_hat = (1j*C00 + C01 - C10 - 1j*C11)/sqrt(2.) #Miracle coin


#Define :
pi=np.pi
cos=np.cos
sin=np.sin
sqrt=np.sqrt
eye=np.eye
roll=np.roll
kron=np.kron
zeros=np.zeros
exp=np.exp
empty=np.empty
outer=np.outer


def Schmidt(l, phi,P,ShiftPlus,ShiftMinus,S_hat,H,X,F,M): #Calc. Schmidt norm to steps N
posn0 = zeros(P)
posn0[m] = 1
psi0 = kron(posn0,(cos(l/2)*coin0 + exp(1j*phi)*sin(l/2)*coin1))
psiN = psi0 #Initialisation

for i in range(1, m + 1):
if (i)%3==0:
psiN = np.linalg.matrix_power(H,1).dot(psiN) #B
else:
psiN = np.linalg.matrix_power(X,1).dot(psiN) #AA
prob1 = empty(P)
prob2 = empty(P)
prob3 = empty(P)
prob = empty(P)
for k in range(P):
posn = zeros(P)
posn[k] = 1
M_hat_k = kron(outer(posn,posn), eye(2))
X_hat_k = kron(outer(posn,posn), C00)
W_hat_k = kron(outer(posn,posn), C11)
proj = M_hat_k.dot(psiN)
proj1 = X_hat_k.dot(psiN)
proj01 = X_hat_k.dot(kron(posn,(coin0)))
proj2 = W_hat_k.dot(psiN)
proj02 = W_hat_k.dot(kron(posn,(coin1)))
prob1[k] = (proj02.dot(proj2.conjugate())*proj01.dot(proj1.conjugate()).conjugate()).real
prob2[k] = (proj02.dot(proj2.conjugate())*proj01.dot(proj1.conjugate()).conjugate()).imag
prob3[k] = ((proj1.dot(proj1.conjugate())).real - (proj2.dot(proj2.conjugate())).real)/2
n1 = n2 = n3 = 0
for j in range(P):
n1 = prob1[j] + n1
n2 = prob2[j] + n2
n3 = prob3[j] + n3
n = sqrt(n1**2 + n2**2 + n3**2)
S = sqrt(0.5 - n) + sqrt(0.5 + n)
Sp = S/(sqrt(2)*2*pi**2)
return Sp

#...Evaluation...:
for m in range(1, N + 1):
P = 2*m+1 # number of positions
ShiftPlus = roll(eye(P), 1, axis=0)
ShiftMinus = roll(eye(P), -1, axis=0)
S_hat = kron(ShiftPlus, C01) + kron(ShiftMinus, C10)
H = S_hat.dot(kron(eye(P), C_hat)) #Hadamard coin plus shift
X = S_hat.dot(kron(eye(P), K_hat)) #Grover coin plus shift
F = S_hat.dot(kron(eye(P), F_hat)) #Fourier coin plus shift
M = S_hat.dot(kron(eye(P), M_hat)) #Miracle coin plus shift
ans, err = integrate.dblquad(lambda l,phi: Schmidt(l, phi,P,ShiftPlus,ShiftMinus,S_hat,H,X,F,M), 0, 2*pi,
lambda x: 0,
lambda x: pi)
DKP_XXH[m-1] = ans
print(DKP_XXH[m-1])

print(DKP_XXH)
\end{verbatim}
\end{scriptsize}}

\end{document}